\documentclass{aa}  
\usepackage{graphicx}
\usepackage{amsmath}
\usepackage[dvips,usenames]{color}
\usepackage{natbib}
\def\ea{{\it et al.} }

\begin{document}
\titlerunning{Near-IR jet of  PKS0521-365 }
\authorrunning{R. Falomo et al  }

\title{The  jet of the BL Lac object  PKS 0521 -365 in the near-IR
: \\ MAD adaptive optics observations. 
   %
  \thanks{ Based on observations collected at ESO, Paranal, Chile,
  as part of MAD Guaranteed Time Observations.
        }
       }
\author{  R. Falomo
	\inst{1}\and
	E. Pian 
	\inst{2}\and
	A. Treves 
	\inst{3}\and 
	G. Giovannini\inst{4,5}, T. Venturi\inst{4}, 
	A. Moretti\inst{1}, C. Arcidiacono\inst{1}, J. Farinato\inst{1}, 
	R. Ragazzoni\inst{1},
          E. Diolaiti\inst{6}\and
           M. Lombini \inst{6}\and
           F. Tavecchio \inst{7}\and
          R. Brast
          \inst{5}\and
          R. Donaldson
          \inst{8}\and
          J. Kolb
          \inst{8}\and
          E. Marchetti
          \inst{8}\and
          S. Tordo
          \inst{8}
          }
 \offprints{renato.falomo@oapd.inaf.it}
 \institute{Osservatorio Astronomico di Padova, INAF,
            vicolo dell'Osservatorio 5, 35122 Padova, Italy 
            \and
                Osservatorio Astronomico di Trieste, INAF, via Tielopo, Trieste, Italy
                \and
                Universita' dell'Insubria (Como), Italy
                \and
                Istituto di Radioastronomia, INAF \ c/o CNR, Via Gobetti 101               
		40129 Bologna, Italy
		 \and
	    Dipartimento di Astronomia - Bologna University - via Ranzani 1 40127 
	    Bologna, Italy
                \and
	     Osservatorio Astronomico di Bologna, INAF,
	     via Ranzani 1, 40127 Bologna, Italy
	     \and
	     Osservatorio Astronomico di Brera, INAF, via Bianchi 46, Merate, Italy
	     \and
	     European Southern Observatory, 
	    Karl-Schwarzschild-Str. 2, 85748 Garching bei M\"unchen, Germany 
}

\date{Received ... ; accepted ...}

\abstract
{BL Lac objects are low--power active nuclei 
exhibiting a variety of peculiar properties that are caused by the presence 
of a relativistic jet and  orientation effects.  } 
{We present here adaptive optics near-IR images at high spatial resolution 
of the nearby BL Lac object PKS 0521-365, which is known to display  
a prominent jet both at radio and optical frequencies. 
}
{The observations were obtained in Ks--band using the ESO  
multi-conjugated adaptive optics demonstrator at the Very Large Telescope. 
This allowed us to obtain images with 0.1 arcsec effective resolution. 
We performed a detailed analysis of the jet and its related 
features from the near-IR images, and combined them with images 
previously obtained with HST in the R band 
and by a re-analysis of VLA radio maps. 
 }
{We find a remarkable similarity in the structure of the 
jet at radio, near-IR, and optical wavelengths.
The broad--band emission of  the jet knots is  
dominated by synchrotron radiation, while the nucleus also exhibits 
 a significant inverse Compton component.
We discovered the near-IR counterpart of the radio hotspot and found 
that the near-IR flux  is 
consistent with being a synchrotron emission  from radio to X-ray. 
The bright red object ({\it red-tip}), detached but 
well aligned with the jet, is well 
resolved in the near-IR and has a linear light profile. Since  it 
has no radio counterpart, we propose that it is a 
background galaxy not associated with the jet. 
}
{The new adaptive optics near-IR images and previous observations at other frequencies 
allow us to study the complex environment 
around the remarkable BL Lac object PKS 0521-365. These data 
exemplify the capabilities of multi conjugate adaptive optics observations 
of extragalactic extended sources. 
}
 
\keywords{galaxies: individual (PKS 0521-365) --
                 Instrumentation: adaptive optics }

\maketitle
%

\section{Introduction}

BL Lac objects are a class of low--power active galactic nuclei (AGN)  characterized by
a lack or extreme weakness of  emission lines observed in all other types of AGN. They 
also exhibit strong and rapid flux variability and significant polarization. The  emission
is dominated by a non--thermal component extending from radio to very high energies. These active
nuclei are hosted by massive elliptical galaxies that appear in most cases unperturbed and located
in moderately rich environments. A widely accepted model that explains these properties is based
on the idea that these nuclei  emit a relativistic jet oriented close to the
line of sight of the observer. A direct consequence of this interpretation is that there are
several other objects intrinsically identical to BL Lacs but with a  misaligned jet that thus
exhibits different nuclear properties. This parent population has been identified with low--power
radio galaxies  \cite[e.g.,][]{Urry95} .

Since the jet is closely aligned with the line of sight, it is very difficult to observe 
unless the angular resolution is high. In the radio band, jet detection is indeed possible and a
large fraction of objects classified as BL Lacs show a signature of a jet,
often with evidence of superluminal motion \cite[e.g.,][]{Homan01,Giroletti04}. 
  On the other hand, the jet is  rarely detected in the optical and X-ray bands. This depends
both on the more limited angular resolution and on the short lifetime of the high energy
electrons  producing the non--thermal emission of the jet.

The first systematic searches for angularly resolved multiwavelength counterparts of
radio jets in AGNs were carried out following the advent of high resolution imaging facilities
such as HST and Chandra. The individual knots within the jets of M87 and 3C273 have been
studied in detail at optical and X-ray frequencies (e.g., Sparks Biretta \& Macchetto 1994;  
Marshall et al. 2001, 2005; Jester et al. 2007), and in M87, evidence of knot
variability has been found in both optical and X-ray bands  \citet{Perlman03,Harris03,Harris06}. Jets of higher redshift radio-loud objects have been  detected and
resolved  with HST and Chandra. About 20 blazars have been found to have optical and X-ray
resolved jets to date (e.g., Scarpa et al 1999; Schwartz et al 2000; Sambruna et al.
2002; 2004; 2007; 2008; Marshall et al. 2005; Tavecchio et al. 2007).
A detailed study of the optical jet based on HST images was completed for: 
PKS 0521-365, PKS 2201+04 and 3C 371 \cite{Scarpa99}. 

Stronger evidence  of a prominent optical jet in a BL Lac object has been found for PKS
 0521-365, and in this paper we present high resolution near-IR images of the source obtained
 at  ESO-VLT. The new images were acquired using an adaptive optics (AO) camera, built to
 probe the new capabilities of the technique, and a re-analysis of the VLA radio
 data allow us to investigate accurately the properties of the jet and the surrounding
 structures.
 
 We adopt the concordance cosmology with
 H$_0$ = 70 km s$^{-1}$ Mpc$^{-1}$, $\Omega_m$ = 0.3, and $\Omega_\Lambda$ = 0.7.


\section{PKS 0521-365}

The radio source PKS 0521--365 is a well-studied nearby object \cite[z=0.0554; ][]{Danziger79}. 
It is one of the most remarkable extragalactic objects of the southern sky, because it
exhibits a variety of nuclear and extra-nuclear phenomena. First classified as an N galaxy,
and then as a BL Lac object, it shows the strong narrow and broad emission lines 
typical of type 1 active nuclei both in optical and UV \citep{Ulrich81,Danziger83,Scarpa95}. 
Interestingly, the source has a prominent radio, optical, and X-ray jet \citep{Danziger79,Keel86,falomo94,Macchetto91,Scarpa99,Birkinshaw02}, 
which resembles that of the nearby radio galaxy M87 \cite{
Sparks94}. Tingay et al. (1996; 1998),  Tingay and Edwards (2002) and 
Giroletti et al (2004) studied the structure of this source on 
sub-pc scales in the radio.

The optical jet was observed with HST by Macchetto et al (1991) and by Scarpa et al (1999), 
who  found a
structure consisting of a bright knot close to the nucleus and a diffuse and knotty
structure that extends out to $\sim$ 6 arcsec from the nucleus. Farther out along the 
jet (but well separated), at $\sim$ 9 arcsec, there is resolved compact emission 
(also called the ``red tip'') \cite{falomo94,Scarpa99}.

The overall energy distribution was discussed by  Pian et al. (1996), who  proposed that the
jet is oriented at about 30 degrees with respect to the line of sight.  

\section{Observations and data reduction}

\begin{figure}[h]
  \centering
\includegraphics[width=.5\textwidth]{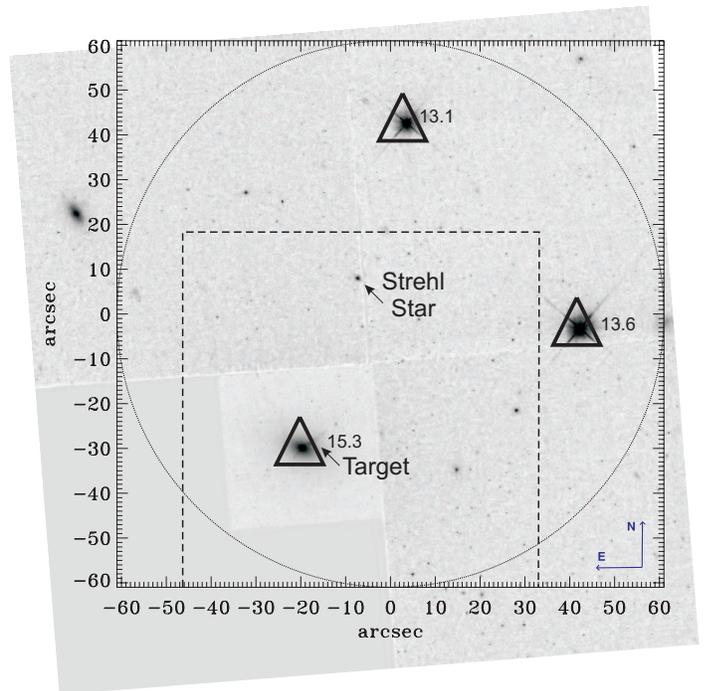}
  \caption{
The 2 arcmin field of view (circle) corrected by the MAD MCAO system. 
The area inside the dashed box was 
imaged by the NIR camera. The greyscale image is a mosaic of HST + WFPC2 images (F702w filter).
The triangles correspond to the LO AO reference stars. R-magnitude is reported for each star. At
the centre of the field the arrow identifies the star used to compute the SR (see Table \ref{madobs} ).
}\label{madfov}
\end{figure}

%
\begin{figure}[ht]
  \centering
 \includegraphics[width=.47\textwidth]{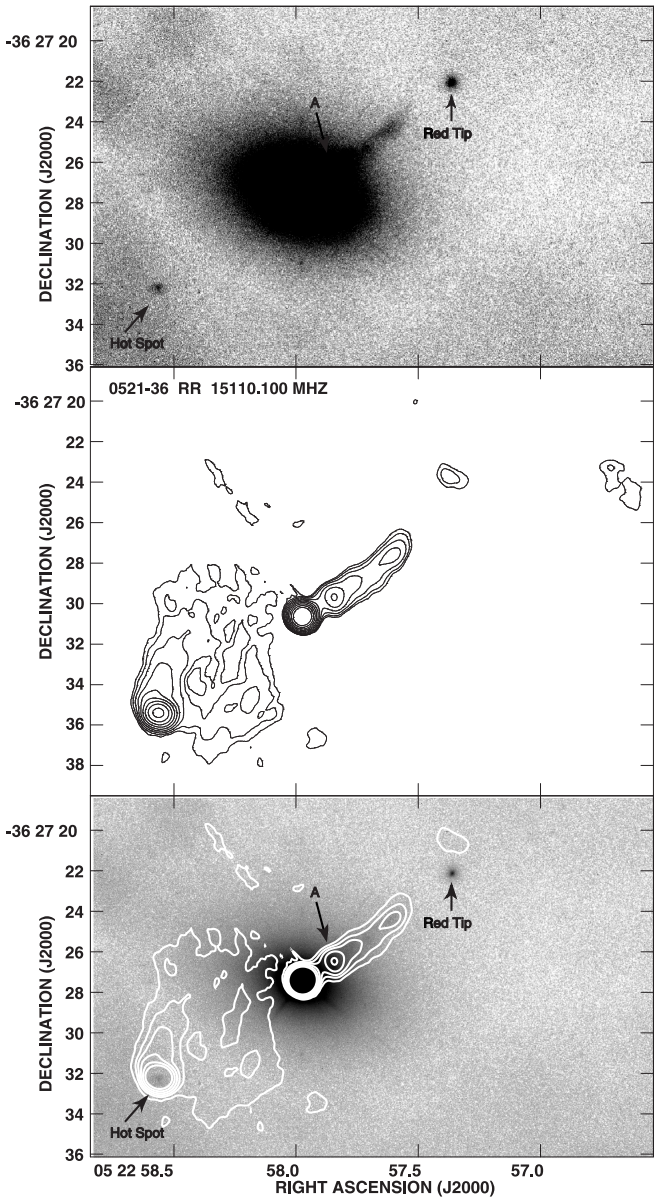}
   \caption{
  Upper panel: The image of PKS 0521--365 observed by MAD in the Ks band. 
Middle panel:  The contours represent the VLA radio map at 15 GHz. 
 Contour levels in the middle panel are: 1,4,8,16,32,64 mJy/beam.
 Bottom panel:  Contour levels in the bottom panel are: -1,1,2,4,8,16,32,64,128,256,512 mJy/beam (0.6x0.6 arcsec)
}\label{fig:ima0521madvla}
\end{figure}

\subsection{MAD near-IR AO images }

Near--infrared images of the  target were obtained using the  ESO Multi conjugate
Adaptive optics Demonstrator (MAD). MAD is an experiment devoted to demonstrating the
feasibility of the Multi Conjugate Adaptive Optics (MCAO) technique  as a test--bench
for the European Extremely Large Telescope.  MAD was mounted onboard the UT-3 Very
Large Telescopes (VLT) and accomplished the first MCAO on--sky observations. 
MAD successfully implemented two MCAO techniques: the star oriented technique based 
on three Shack-Hartmann wavefront sensors and the layer oriented technique (LO) 
\cite{Ragazzoni00a,Ragazzoni00b}  based on a multi pyramid sensor \cite{rag96}. Both
devices can use reference stars on a 2 arcmin technical Field of View; in particular, in the
LO mode up to 8 pyramids can be positioned where the focal plane images of the
reference stars form, splitting the light into four beams. A beam splitter illuminates two
identical objectives conjugated at different atmospheric altitudes. These optics coadd
the pupil images of all reference stars on two fast readout CCDs simultaneously. 
These measurements are the input for  computing the corrections 
to be applied to the two deformable mirrors, which are  conjugated to 
zero and 8.5 km altitude.
The observations presented here were obtained using the LO wavefront sensor only.


The images were obtained in a MCAO configuration with  both deformable mirrors, and 
used two reference stars (R-band magnitudes: 13.6, 13.0 ) and the core of the target (R = 15.3) 
to perform wavefront sensing (see Fig. \ref{madfov}). 
These correspond to an overall integrated magnitude of R $\sim$ 12.5.
Given that MAD wavefront sensing CCD
sensitivity  peaks in the V band, we  estimate an overall V $\sim$ 12 
integrated magnitude using  colours of stars available from USNO. 

In contrast to MCAO observations carried out in crowded regions, 
such as a Globular Cluster (\cite{Moretti09, Bono09}, 
this observation demonstrates that the case where MCAO allows not only 
for a uniform and yet higher Strehl Ratio,
but  makes  observations possible at about diffraction limit resolution, 
because of the increasing of the AO sky coverage. 
In fact, given the distance of the neighbourhoods stars from the target 
any Single Conjugated AO observations
could use the object itself as a reference source. 
In this MCAO case the gain in terms of
useful flux to achieve the AO compensation is more than an order of magnitude. 

The observations were carried out with the telescope  pointing close to the zenith, 
with airmass 1.05, and the ESO DIMM
seeing monitor measured an average seeing in V band of 1.1 $\pm 0.1$ arcsec .
We computed the FWHM and the strehl ratio (SR) (see Table \ref{madobs} ), 
using a relatively bright 
star located near the centre of the field of view (see Fig. \ref{madfov} ).


Images of the  target were obtained following a 5--position jitter pattern,
with offsets of 5 arcsec, to ensure an adequate subtraction of the sky
background. The entire data set consists of 20 exposures with slightly different
positions with the pattern, for a total exposure time of 3600 s.


Each individual image was trimmed and flat fielded with the appropriate flat--field frame
derived from several images obtained on the sky at the beginning of the night in the $K_s$
filter. A bad--pixel mask was created by analysing the ratio of two flat--field images with
significant different exposure levels  to identify the pixels that 
not responded adequately  to the signal intensity. 

For each observation, we constructed a reference sky image, which was subtracted from the 
science
frame  to obtain the net intensity frame. This is particularly important for NIR
observations since they are dominated by the sky emission. The  sky
variation was $\sim 2$\%.
This allowed us to construct a sky-background frame from the median of all science frames
gathered for each target. The
final background frame was normalized to the median counts of each scientific frame
before being subtracted.

Finally, we combined all sky--subtracted and registered frames into a final image. 
In Table \ref{madobs},  we list
the total exposure time of the images that have been effectively used to construct the final
image.  As a first measure of the quality of the photometry, we also list the average FWHM
measured in each combined frame.

\begin{table}[htbp]
\centering
\caption{Journal of MAD observations }
\label{madobs}
\begin{tabular}{llcccc}
\hline\hline
Dataset & Date & Seeing & T$_{exp}$ &FWHM  & SR \\
\hline    
 a)    & 27/9/2007  &  0.99  & 900 &  0.11 & 14.4 \\        
 b)    & 27/9/2007  &  1.02  & 900 &  0.12 & 26.3 \\  
 c)    & 27/9/2007  &  1.11  & 900 &  0.09 & 22.0 \\  
 d)    & 27/9/2007  &  1.05  & 900 &  0.10 & 15.9 \\  
 All   & 27/9/2007  &  1.04  & 3600 &  0.12 & 17.6 \\  
\hline    
\end{tabular}
\end{table}


\subsection{HST optical images}

PKS 0521-365 was observed with HST using WFPC2 in the R (F702W) filter as part of 
the HST snapshot survey of BL Lac objects \cite{Scarpa00,Urry00}. 
The target was imaged with the PC device for a total exposure time of 305 sec. 
These images were discussed in detail in \cite{Scarpa99} and are used here 
to compare with NIR data obtained with MAD at VLT.

\subsection{ VLA radio data and morphology}

VLA radio observations at 1.4 and 15 GHz were retrieved from archive data. Both
observations were taken with the array in the BnA configuration. The 1.4 GHz observations
were gathered on July 1987 (Project ID AV151) and those at 15 GHz on March 1985 (Project ID
AL508). PKS 0521--365 was observed for only 2 minutes at 1.4 GHz, while the 15 GHz observations
had a total time on source of 2.5 hours.

Published images are available at both frequencies (1.4 GHz Keel, 1984; 15 GHz,
Keel et al. 1986), although to take advantage of today's more powerful
computer and data reduction facilities, we re--analysed both data sets. To obtain
our final images, we calibrated the data with the present standard procedures in
AIPS,  accurately edited the visibilities, and carried out a few  self--
calibration cycles: i.e., 4 in phase only for 1.4 GHz data and 2 in phase and one more
in phase and gain for 15 GHz data. The final images were produced with the IMAGR
task. The parameters of the  images discussed here are given in Table \ref{tab:radio}.


In Fig. \ref{fig:ima0521madvla}, 
we show a full resolution (HPBW = 0.3 arcsec) 15 GHz
image with radio contours overlaid on the near-IR image. The radio source is dominated
by a bright core with a peak flux density of 2.6 Jy. The well known one--sided jet 
\cite[e.g.][]{Keel86} is clearly visible on the N--W side, while on the opposite side a bright hot spot
is present. No indication of a hot spot is visible on the side of the main radio jet. At
lower resolution (Fig. 2), the W lobe is marginally visible with faint substructures. The
eastern lobe has a higher brightness and exhibits a morphology typical of FR II sources, where
the lobe fills the region  between the core and the hot spot. The lobe structure is in
agreement with a backflow of the radio plasma after a strong interaction between the hot
spot at the end of the relativistic jet and the surrounding medium.

%
\begin{table}[htbp]
\caption{Parameters of the radio images }
\label{tab:radio}
\centering
\begin{tabular}{lcc}
\hline\hline
Frequency & HPBW & r.m.s. \\
 {GHz} & {`` (degree)} & {mJy/beam}\\
      &                 &    \\
1.47  & 4.11 x 1.21 (-7) & 0.6  \\
15.11 & 0.6 x 0.6        & 0.25 \\
15.11 & 0.3 x 0.3        & 0.28 \\
\hline\hline 
\multicolumn{3}{l}{Col. 1: Observing frequency. }\\
\multicolumn{3}{l}{Col. 2: HPBW and Position Angle of the major axis in degree.}\\
\multicolumn{3}{l}{Col. 3: 1 rms level}\\
\end{tabular} 
\end{table}


\section{Results }

In Fig. \ref{fig:ima0521madvla}, we show the near-IR image of PKS 0521-365
obtained by MAD compared with the radio map at 15 GHz. At both near-IR and
radio wavelengths, the jet is clearly well detected together with the  nucleus
and the bright knots. On the opposite side of the jet, there is  fainter
emission within the MAD image which is coincident with the radio hot spot.

\begin{figure}
  \centering
    \includegraphics[width=.40\textwidth]{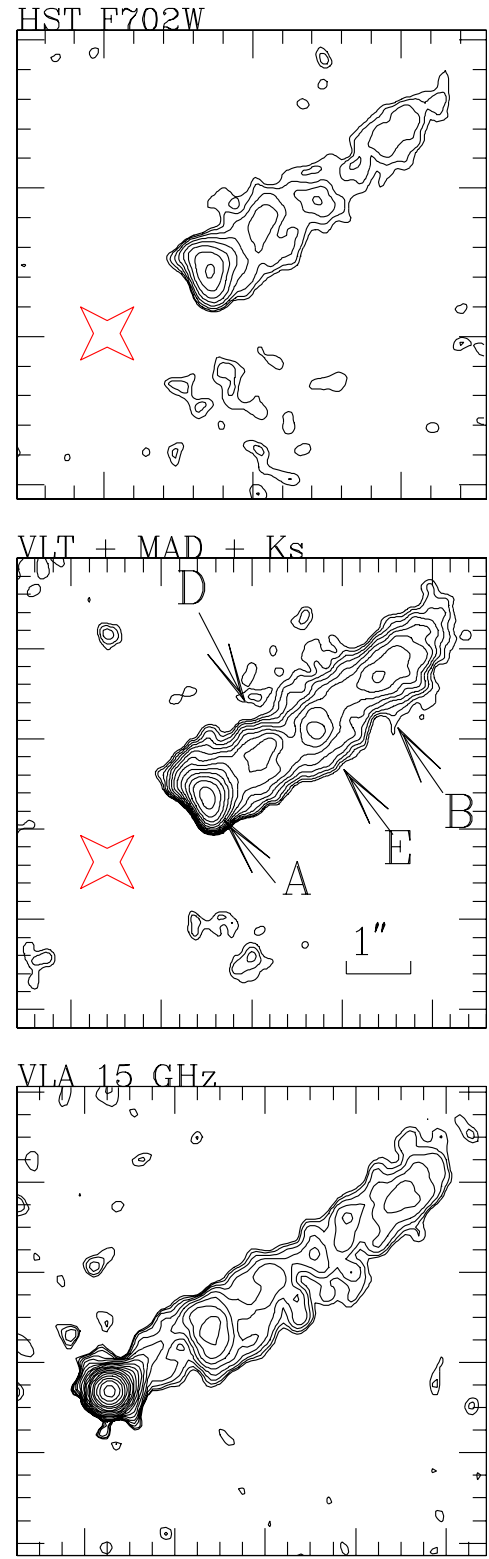}
  \caption{Upper panel. 
The contour plot of the jet of PKS 0521--365 observed by MAD in the Ks band (
middle panel) compared with the image of the jet observed in  R band  by HST+
WFPC2 (top panel) and the radio map at 15 GHz obtained by VLA (bottom panel).
The large star represents the position of the (subtracted) nucleus in both
optical and near-IR bands. In the bottom panel,  the radio jet is shown at the
same angular resolution as the near-IR data. The nuclear source has not been
subtracted. A collimated jet with the same position angle as the VLBI pc-scale
jet is present near the core. The jet becomes resolved
    transversally at less than 1" from the core. 
}\label{jet0521combined}
\end{figure}

\subsection{ Analysis and measurements of the jet }

\begin{figure*}[ht]
  \centering
\includegraphics[width=.7\textwidth]{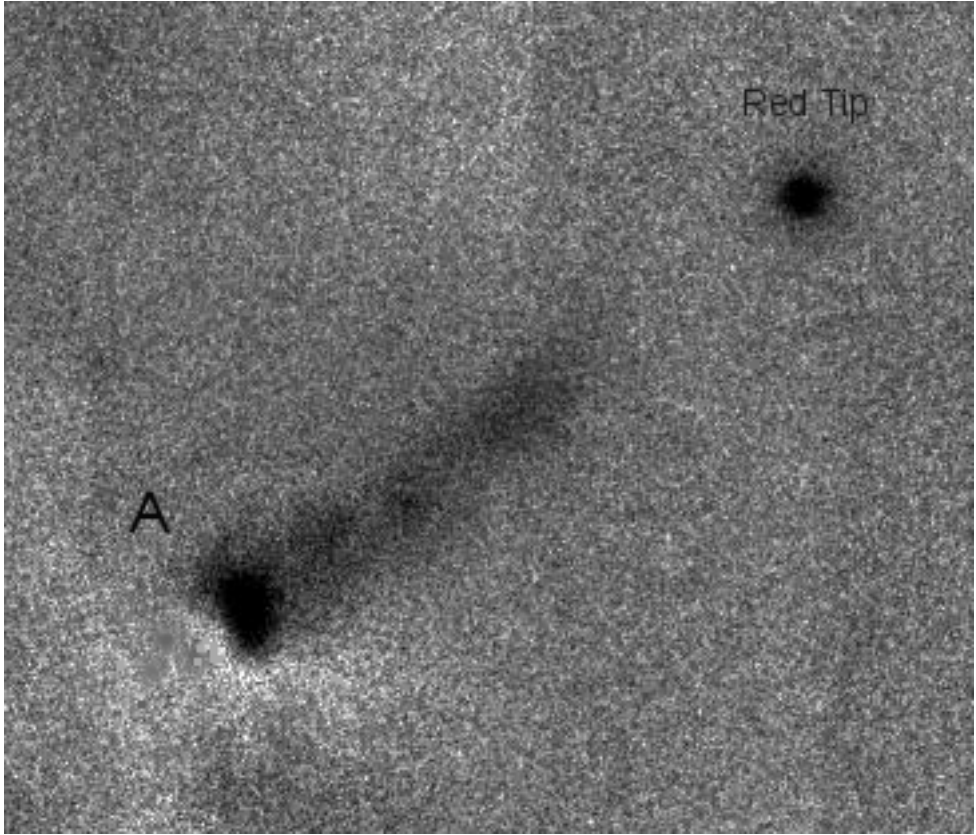}
  \caption{
The near-IR image of the jet of the BL Lac object 
PKS 0521-365,  as observed by MAD at VLT in Ks filter. 
The host galaxy and the nucleus have been subtracted from the original image 
(see text). The field of view shown is 10 arcsec.
}\label{jet0521}
\end{figure*}

\begin{figure}
   \centering
 \includegraphics[width=.5\textwidth]{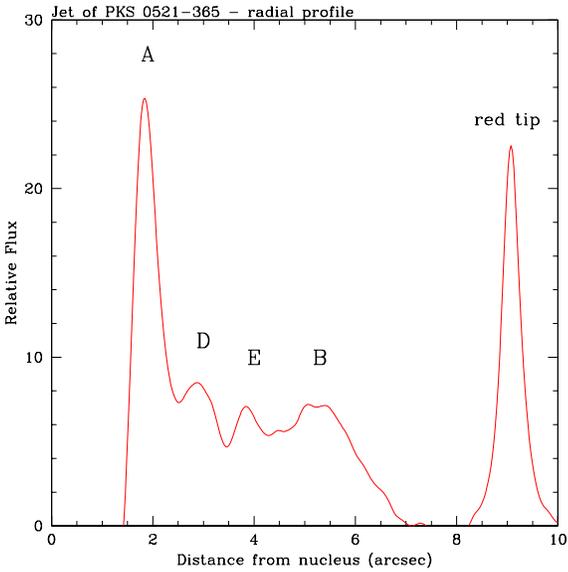}
   \caption{
 The brightness profile along the jet of the BL Lac object 
 PKS 0521-365  and of the red-tip as observed by MAD at VLT in Ks filter. 
 The profile was obtained by integrating the signal in a strip of fixed size of 
 0.56 arcsec width between the nucleus and the red-tip. 
 The main peaks correspond to the knots labelled  in Fig. \ref{jet0521combined}. 
 The flux from both the nucleus and  the host galaxy has been removed.
 }\label{jet0521prof}
\end{figure}

To enhance  the structure of the jet in the near-IR image, we have
subtracted the emission of both the nucleus and the host galaxy from the
original image. To subtract the host galaxy, we performed a detailed 
two--dimensional surface photometry analysis of the source and derive a clean
model of the object. The adopted procedure  is very similar to that
described in \cite{falomo00}. The emission from the jet after the subtraction
of the nucleus and the host galaxy is shown in Fig. \ref{jet0521}. The
bright knot A at 1.9 arcsec from the nucleus and the diffuse emission from
the jet are clearly detected in near-IR as well as the resolved feature in
the red-tip. In Fig. \ref{jet0521prof}, we show  the brightness profile
along the jet.

To compare the features in NIR, 
optical, and radio bands, we  applied the same procedure used for the 
MAD image to subtract the host galaxy and the nucleus from the HST R band image.
In Fig. \ref{jet0521combined} we compare images of the jet in the three different bands.
The  overall morphology of the jet in the near-IR is similar to that  
 observed in the optical \cite{Scarpa00} and  at 2cm \cite{Keel86}.  
The general shape and  total length of the radio jet
are almost exactly reproduced in the near-IR, in the 
optical, and at radio frequencies. 

Following the previous analysis of the jet, 4 knots  can be easily 
recognized. These have been labelled in Fig. \ref{jet0521combined},  
according the definitions introduced in  \cite{falomo94}. 
To construct the spectral emission of these components of the jet, we 
measured the flux of the four knots by adopting a fixed square aperture of 1 arcsec.
In Table \ref{tab:jet}, we report the positions and the measurements of the knots in all bands.  
In the radio band, the knots are well resolved with sizes given in Table \ref{tab:jet}. 


%
\begin{table*}[htbp]
\centering
\caption{Jet measurements of PKS 0521-365}
\label{tab:jet}
\begin{tabular}{cccccc}
\hline\hline
Feature  &  $K_s^a$         & $R^b$           & $F_{Radio}^c$    & Size$^d$   & $F^e_{X-ray}$                    \\
         &  (mag)           & (mag)           & (mJy)            &  (arcsec)  & (mJy)                            \\
\hline
Nucleus  &  $11.9 \pm 0.05$ & $15.30 \pm 0.05$ & $2600 \pm 100$  & ... & ($6.8 \pm 0.1$) $\times 10^{-4}$ \\
A        &  $17.6 \pm 0.1$  & $20.3  \pm 0.1$  & $58 \pm 5$      & 0.7x0.5  & ($14 \pm 3$) $\times 10^{-6}$    \\
D        &  $18.4 \pm 0.2$  & $21.6  \pm 0.2$  & $20 \pm 3$      & ...    & ...                              \\
E        &  $18.5 \pm 0.2$  & $22.1  \pm 0.2$  & $14 \pm 2$      & ...    & ...                              \\
B        &  $18.5 \pm 0.2$  & $22.0  \pm 0.2$  & $19 \pm 3$      & 1.3x0.6 & ...                              \\
red-tip  &  $17.9 \pm 0.2$  & $21.3  \pm 0.2$  & $ ... $         &  0.6x0.4 & ...                              \\
hotspot  &  $19.9 \pm 0.2$  & $ > 22$          & $588 \pm 6$     &  ...    & $\approx 4 \times 10^{-7}$       \\
\hline\hline   
\multicolumn{6}{l}{$^a$ $K_s$-band magnitude from VLT MAD measurements} \\
\multicolumn{6}{l}{$^b$ $R$-band magnitude from HST WFPC2 measurements in F702W filter, 
not corrected for Galactic reddening} \\
\multicolumn{6}{l}{$^c$ Radio measurements at 15 GHz from VLA high resolution map}\\
\multicolumn{6}{l}{$^d$ Measured on the radio maps}\\
\multicolumn{6}{l}{$^e$ X-ray measurements at 1 keV from Chandra images, \ 
	corrected for Galactic absorption (Birkinshaw et al. 2002)} \\
\end{tabular}
\end{table*}

\subsection{The nature of the red tip} 		

The near-IR image obtained by MAD also clearly detects and resolves the diffuse emission at about 9
arcsec NW from the nucleus (see Table 3 for details). The observed apparent mag in K = 17.9. 
The object is very red: R- K = 3.4 (using  R mag = 21.3 derived from HST image). 
This feature is well resolved in the near-IR 
and has an average surface brightness that is 
well described by a linear (disk like) profile.

We performed a detailed analysis of the radio emission in the region around the red tip . This revealed
 extended radio emission at about (RA=05 22 57.38  ; Dec= -36 27 23.8 ) that is not exactly
coincident with the optical position (RA=05 22 57.3 ; Dec= -36 27 25.5) of the red tip but
sufficiently close to be considered physically associated. The shape of this radio structure suggests
that it has the characteristic morphology of a head-tail radio galaxy. 
Based on these coincidences we
argue that the red tip could be a background, faint radio--galaxy.

\subsection{ The near-IR counterpart of the radio hot spot } 

\begin{figure}
 \centering
 \includegraphics[angle=0,width=.45\textwidth]{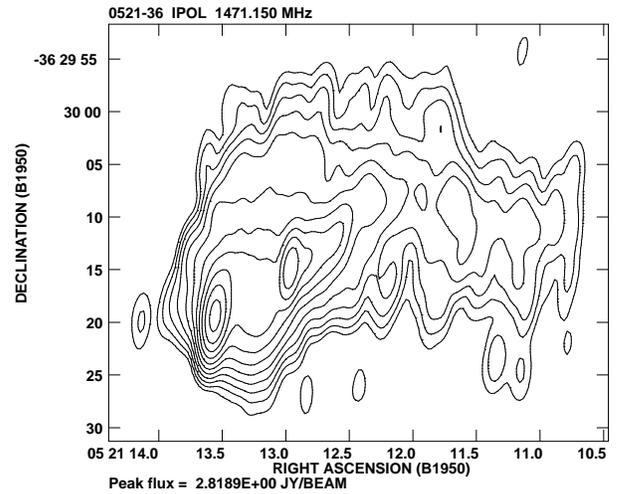}
 \caption{ The radio map at 1.4 GHz of PKS 0521-365. Contour levels are: 
 -2,2,4,8,16,32,64,...,2048 mJy/b. The beam size is 4.11 x 1.21 (arcsec); p.a. -7} 
\label{radio14ghz}
\end{figure}

In the 1.4 GHz image (see Fig. \ref{radio14ghz}; 
HPBW = 4.11" x 1.21" in PA -7), the hot
spot is the dominant feature (peak flux = 2.8 Jy/beam) because of its steep spectrum, while
the core is fainter (1.8 Jy) with respect to the 15 GHz flux density, which is self--absorbed.
This image confirms the lack of a hot spot like structure on the jet side. At 1.4 GHz, the
radio emission is extended westwards, with a low brightness morphology typical of FR I
sources, while the East lobe shows a large extension in N-S but always in the back-flow
region in--between the core and the hot spot. No indication of a counter--jet is visible. 

In our near-IR image we searched for emission that is 
coincident with the radio hot spot and found a faint (K$_S$ $\sim$ 20) compact feature at the position 
RA = 05:22:58.54  DEC=-36:27:35.67  (Equinox: 2000.0).
This is coincident with the peak emission of the hot spot to within 0.2 arcsec. 
At this position, there is no optical counterpart in the  R--band HST image. 
We estimate a magnitude limit of R $>$ 22 for this feature. In a few other cases, the 
optical-NIR counterpart of radio hot-spots was detected 
\cite[e.g.][]{Dreher86,Hartman99,Meisenheimer97,Hardcastle01,Prieto03}.

\section{Discussion }
\subsection{Spectral energy distribution of jet components}

\begin{figure}
  \centering
  \includegraphics[width=.5\textwidth]{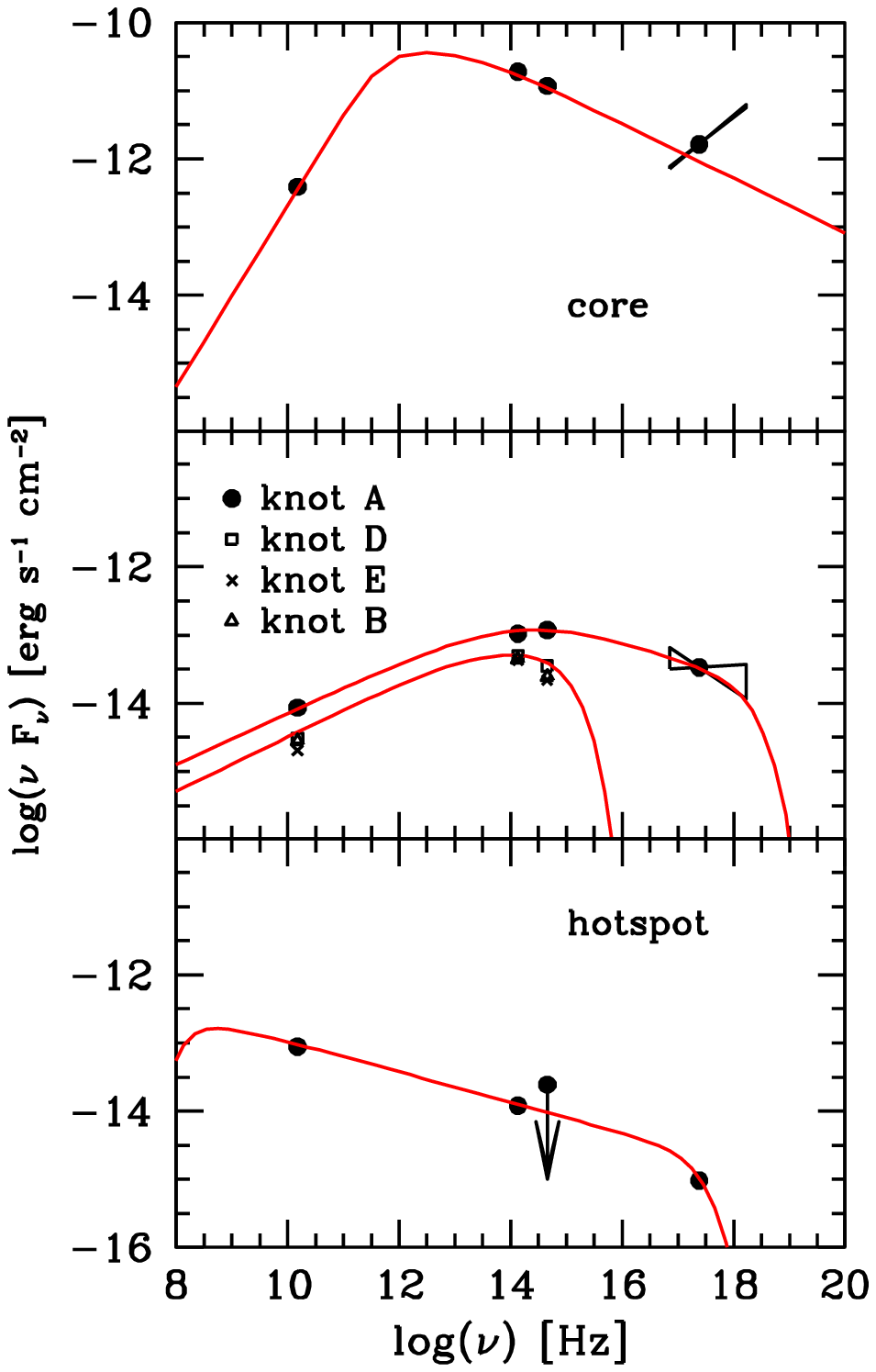}
  \caption{
  The spectral energy distributions of the different components of the BL Lac object 
  PKS  ~0521-365 ($z = 0.055$) :  nucleus (top), jet knots A,D,E,B (middle) and hot spot (bottom).  
  All X-ray data are from Birkinshaw, Worrall, \&  Hardcastle (2002), 
  and are corrected for the absorption of Galactic neutral hydrogen
  with a column density of $3.37 \times 10^{21}$ cm$^{-2}$. 
  The optical data of the nucleus and of the jet knot are
   measured from the HST image of Scarpa et al. (1999). The radio data of the
  nucleus are from VLA and VLBI measurements (se text). 
  The near-IR data  have been measured from our MAD images. 
}\label{sed0521}
\end{figure}

We have combined our near-IR measurements of the nucleus, jet knots, and hotspot with
the available optical, radio and X-ray fluxes  to construct overall 
spectral energy distributions (SED) of main components. 
The optical
and near-IR fluxes were corrected for Galactic dust absorption using $E_{B-V} = 0.039$
(Schlegel et al. 1998), and the \cite{Cardelli89} extinction curve. 
The X-ray fluxes \cite{Birkinshaw02} were corrected for Galactic absorption assuming 
$N_H = 3.37 \times 10^{20}$ cm$^{-2}$. 
The resulting  spectral energy distributions are reported in Fig. \ref{sed0521}.  

The SED of the core has been compared (assuming modest multiwavelength variability) with a
classical synchrotron emission model in a single region of radius $R_0$ filled
homogeneously with a magnetic field B, in equipartition with the relativistic particles,
following the formalism of Ghisellini et al. (1998) and Sari, Piran, \& Narayan (1998). In
the emitting region, the electrons are accelerated by a relativistic shock to a power--law
distribution with a minimum Lorentz factor $\gamma_m$:
$N(\gamma_e) \propto \gamma_e^{-p} d\gamma_e$, with $\gamma_e \ge \gamma_m$.  The spectral
power peaks at the cooling  frequency $\nu_c$, corresponding to the electron energy
$\gamma_c$ above which the electron radiative losses become significant (note that in
Ghisellini et al. (1998) this is named $\gamma_{min}$).  The spectral index $\alpha_0$ below
  $\nu_c$ ($F_\nu \propto \nu^{-\alpha}$) is related to the electron distribution index
by $p = 2 \alpha_0 + 1$. The spectrum above $\nu_c$ has an index
$\alpha_1 = \alpha_0 + 0.5$. In the case of the jet knots and hotspot, we found that this
simple  synchrotron model is not satisfactory for the description of the SED, therefore 
 we adopted a
curved electron distribution similar to that used in Tavecchio et al. (2002).  In this
formulation, the break random Lorentz factor $\gamma_b$ corresponds to 
$\gamma_c$ in the above simpler model.

In spite of the modest coverage of the points, most of the model
parameters are  well constrained by the data.  First of all, the radio VLA and VLBI
data suggest that the kinematic conditions in the jet of PKS~0521-365 are not extremely
relativistic (Pian et al. 1996; Tingay \& Edwards 2002; Giroletti et al. 2004), which
implies a moderate bulk Lorentz factor for the plasma motion in the jet (see also
Birkinshaw et al.  2002). Second, the fact that an arcsec-scale jet is observed at all
frequencies with a projected size of $\sim$ 7 kpc indicates that our viewing 
angle is not small, so that the light aberration is not relevant.  
As proposed by  Pian et al. (1996), we assume $\Gamma$ = 1.2 for all jet components 
(for the hotspot we assume $\Gamma$ = 1.1, to account for moderate plasma braking at that distance from the jet apex)  and a
jet viewing angle of 30 degrees. These parameters imply a relativistic Doppler factor
$\delta = 1.6$.

\bigskip
{\it The jet core} 
\bigskip

For the nucleus,  the size of the emitting region is constrained by the observed minimum
variability timescale of about 1 day.  The cooling frequency is probably below the
infrared frequencies, which we set to be $\nu \sim 10^{14}$ Hz.  The IR/optical spectrum has a
slope $\alpha_1 = 1.4$ (see also Pian et al.  1996), from which we derive $\alpha_0 =
0.9$.  By assuming a magnetic field and electron density  similar to those reported in
Pian et al. (1996) - which appears reasonable  considering the lack of substantial long-
term multiwavelength variability - we obtain a good fit to the spectrum up to X-ray
energies. However, the model does not reproduce satisfactorily the Chandra spectrum (see
Fig. \ref{sed0521}): this is clearly due to an extra component, which we identify with
inverse Compton scattering of relativistic electrons off the synchrotron photons or off
external radiation fields.  A bright inverse Compton component is known to be present in
this source and to have its maximum power at MeV-GeV energies 
\citep{Lin95,Pian96,Hartman99,Tavecchio}.

The  ratio of the inverse Compton to synchrotron luminosities, parameterized by the
total radiation density $U_{rad}^{\prime}$ in the frame of the plasma and by the magnetic
energy density $U_B$, respectively, i.e., $L_{IC}/L_{syn} \simeq U_{rad}/U_B$ is about 2 for a
magnetic field B = 30 G (see Table \ref{tab:jet}), consistent with the multiwavelength
spectrum (Pian et al. 1996). In our estimate, which imposes a strong constraint on the
magnetic field, we have included the radiation density of the synchrotron photons, of the
broad line region (BLR),  and of the inner accretion disk, i.e., $U_{int}^{\prime}$,
$U_{BLR}^{\prime}$, and $U_{BLR}^{\prime}$, respectively. 
The luminosities and radii of both the BLR and disk   are taken from \cite{Pian96}.  
As previously argued by Pian et al. 1996, the largest contribution to 
the inverse Compton emission is due to SSC, the BLR and accretion--disk 
seed photons contributing only  a small fraction (in total less than 10\%).

\bigskip
{\it The jet and the hotspot} 
\bigskip

For the jet knots and hotspot, which are  resolved in the radio, optical and near-IR images, we
assume an emission region radius of 0.5 kpc. Since we do not detect any evidence of an inverse
Compton component in the X-ray emission of knot A and hotspot (see also Birkinshaw et al.
2002), we have determined the magnetic field  imposing that the inverse Compton
luminosity is lower than 10\% of the synchrotron luminosity.  
The electron density 
is much lower than that in the nucleus (see Table \ref{tab:sed}), 
which is consistent with the presumably
more rarefied medium in the external regions of the jet with respect to the core.

For knots D, E and B we can study the SED using only the radio, 
near-IR, and optical fluxes. 
The lack of X-ray fluxes thus makes the modelling of their SEDs  more
uncertain.  A curved synchrotron model, with a negligible inverse Compton contribution,
accounts well for the data, and in particular for the spectral curvature above the optical
frequencies (that cannot be formally reconciled with the $\Delta\alpha = 0.5$ expected
from simple homogeneous synchrotron radiation) and 
with the  lack of X-ray detection (Birkinshaw et al. 2002). 
In Table \ref{tab:sed}, we  report only the model parameters for
knot D, those of knots B and E being very similar.

\bigskip
{\it Comparison between the SED of the core and knot A} 
\bigskip

As appears from Fig. \ref{sed0521}, the multiwavelength spectrum of the nucleus  differs markedly from
that of knot A.  In particular, the X-ray spectral shapes differ clearly :  that of knot A is consistent
with being the high energy portion of the synchrotron component that reproduces  the radio-to-optical
emission, while the X-ray spectrum of the nucleus is caused by an extra emission component  that we
identify with inverse Compton scattering of the relativistic particles that produce the radio-to-infrared
synchrotron spectrum off synchrotron and/or external photons.  This difference, which is mainly related
to the higher ($\sim$10) cooling frequency of knot A with respect to the jet apex. The density of  the
synchrotron photons in the core is much higher than in knot A (the total luminosity of knot A is a factor
100 less than in the nucleus and the emission region size is at least 4  orders of magnitude larger).
This causes  a more efficient electron cooling in the nuclear region via synchrotron self-Compton. The
spectrum of PKS0521-365 exhibits prominent broad optical and UV emission lines \cite[e.g.][]{Scarpa95}
and therefore the density of the photon--field external to the jet within the BLR radius (a fraction of
a parsec)  is not negligible, while it is irrelevant at the distance of knot A (few kpc). 
Although  electron cooling via inverse Compton off the external photons in the nucleus is
not dominant with respect to synchrotron self--Compton,  it contributes to making the cooling more
efficient \cite[see][]{Ghisellini98}. On the other hand, knot A is located beyond the BLR 
(its distance from the nucleus is about 1-2
kpc; well beyond the BLR region), so that its emitting electrons do
not cool as efficiently as in the nucleus, and the synchrotron component extends up to the
X-rays, with a far less significant contribution from inverse Compton scattering.

 \begin{table*}[htbp]
 \caption{SED modelling of  PKS 0521-365 components}
 \label{tab:sed}
 \centering
 \begin{tabular}{cccccc}
\hline\hline
 %
           & $\alpha_0^a$ & $R_0^b$              & $n_0^c$              & $B^d$                &  $\nu_c^e$            \\
           & 	        &	                   &	                  &	                 &	                 \\
 nucleus   &   0.9        & $1.5 \times 10^{-6}$ & 95000                & 30                   & $1.0 \times 10^{13}$  \\
  knot A   &   0.7        & 0.5                  & $1.3 \times 10^{-4}$ & $9.5 \times 10^{-5}$ & $1.0 \times 10^{14}$  \\
  knot D   &   0.7        & 0.5                  & $6.2 \times 10^{-5}$ & $6.9 \times 10^{-5}$ & $5.8 \times 10^{13}$  \\
  hotspot  &   0.8        & 0.5                  & $5.8 \times 10^{-3}$ & $5.3 \times 10^{-4}$ & $3.0 \times 10^8$  \\
 \hline\hline  	  
\multicolumn{6}{l}{$^a$ spectral index below $\nu_c$.} \\
\multicolumn{6}{l}{$^b$ size of emitting region in kpc.} \\
\multicolumn{6}{l}{$^c$ plasma density, i.e. number of particles per cm$^3$.}  \\
\multicolumn{6}{l}{$^d$ magnetic field in Gauss. } \\
\multicolumn{6}{l}{$^e$ cooling frequency in Hz.} \\
\end{tabular}
\end{table*}

\subsection{Radio morphology of the source}


A comparison between
the VLBI data (see Sect. 5.1) and the present images confirms that the source is 
oriented at a relatively large angle  
with respect to the line of sight and that beaming is not playing a major
role. This is suggested by the good agreement between the jet 
orientation on both the
parsec and the kpc scale, and by the extended structure visible at low
frequency. The 15 GHz image shows that the arcsecond jet is still clearly 
one--sided. From the jet brightness ratio in the full resolution 15 GHz image, we infer
that the jet is still relativistic on the arcsecond scale, with a velocity $>$
0.79c in the inner kpc region and $>$ 0.3c at 5--6 kpc from the core.

Assuming an average jet velocity $\sim$ 0.5c and a constant orientation with
respect to the line of sight, we can derive the expected ratio of the
approaching to the  receding size. The length of the approaching jet should  be at
least 3 times longer than the receding one. This could explain the lack of a hot
spot structure on the jet side. The hot spot should be almost at the end of the
radio structure visible at 1.4 GHz. At this distance, the ISM might have a lower
density and therefore the ISM--jet interaction is not expected to produce 
this structure. We further note that the kpc scale jet is clearly defined and
stops sharply well within the more extended lobe visible at 15 and at 1.4 GHz
extent, and this morphology cannot be created by sensitivity problems. 

A possible explanation for the difference in  morphology between 1.4 GHz and 
15 GHz is a reactivation of 
 this source. According to this interpretation, the visible
jet is due to the more recent activity, while the more extended structure is
related to an older radio activity. In this scenario the eastern hot spot 
would no longer be  connected to the shorter recent counter-jet (not visible because
of relativistic effects), and the western hot spot could be missing due to the
older source age in this region.

\bigskip

\section{Summary and Conclusions}

We have presented high resolution near-IR images of the jet of the nearly BL Lac object 
PKS 0521-365 using an innovative adaptive--optics device (MAD) 
built as a demonstrator for 
multi--conjugated AO imaging. These new data, together with a 
re-analysis of previous 
optical (by HST) and radio (by VLA) data, have provided us with insight into 
a number of 
remarkable features that are associated with this nearby extragalactic source. 
The main results from this study are: 
i) the morphology of the jet is very similar at radio, near-IR, and optical 
frequencies; ii) the emission from the jet knots is dominated by the 
synchrotron component,  
while in the nucleus a significant inverse Compton component is present at high energies; 
iii) we discovered the near-IR counterpart of the radio hotspot and found 
its flux  to be  
consistent with the synchrotron emission from radio to X-ray; 
iv) the resolved red object aligned with the optical jet has 
no radio counterpart and  is thus 
likely to be a background galaxy that is not associated with the jet. 

There are two observations that may  clarify the nature of this peculiar
system. A spectrum of the red tip appears feasible and  may 
confirm that it is a galaxy. Since the distance of the object is small, proper
motion of the order of 0.1 mas of the jet structures are expected, and
in principle should be detectable. 

In addition to the specific results for the target, these observations  
exemplify the capabilities of MCAO observations of extragalactic extended sources.


\begin{acknowledgements}

EP acknowledges support from the Italian Space Agency and the 
National Institute of Astrophysics INAF through grants 
ASI-INAF I/023/05/0 and ASI I/088/06/0. We wish to thank P. Amico
for her support during the MAD observations and the 
ESO staff at Paranal for kind assistance.
\end{acknowledgements}

%

\end{document}